\documentclass[twocolumn,showpacs,preprintnumbers, 
               superscriptaddress,nofootinbib,prl]{revtex4-1}
\usepackage{graphicx, fancybox}
\usepackage{amsmath,amssymb,bm}

\usepackage[T1]{fontenc} 

\usepackage[table]{xcolor}

\begin{document}

\title{Correlations of Energy-Momentum Tensor via Gradient Flow\\
 in  SU(3) Yang-Mills Theory at Finite Temperature 
}

\author{Masakiyo Kitazawa}
\email{kitazawa@phys.sci.osaka-u.ac.jp}
\affiliation{Department of Physics, Osaka University, 
  Toyonaka, Osaka 560-0043, Japan}
\affiliation{
  J-PARC Branch, KEK Theory Center,
  Institute of Particle and Nuclear Studies, KEK,
  203-1, Shirakata, Tokai, Ibaraki, 319-1106, Japan }

\author{Takumi Iritani}
\affiliation{Theoretical Research Division, Nishina Center, RIKEN, Wako 351-0198, Japan}

\author{Masayuki Asakawa}
\affiliation{Department of Physics, Osaka University, 
  Toyonaka, Osaka 560-0043, Japan}

\author{Tetsuo Hatsuda}
\affiliation{Theoretical Research Division, Nishina Center, RIKEN, Wako 351-0198, Japan}
\affiliation{iTHEMS Program and iTHES Research Group, RIKEN, Wako
        351-0198, Japan.}
\begin{abstract}
Euclidean two-point correlators of the 
energy-momentum tensor (EMT) in SU(3) 
gauge theory on the lattice are studied 
on the basis of the Yang-Mills gradient flow.
The entropy density and the specific heat obtained from 
the two-point correlators are shown to be in good agreement with those 
from the one-point functions of EMT.
These results constitute a
first step toward the first principle simulations of
the transport coefficients with the gradient flow.
\end{abstract}
\preprint{J-PARC-TH-0097, QHP-317} 

\maketitle

Various thermal and transport properties of quantum field theories
are encoded in the correlations of  energy-momentum tensor (EMT)
at finite temperature ($T$).  In particular, fluctuation and transport 
properties of hot QCD (Quantum ChromoDynamics) matter  
at finite $T$ have attracted a lot of attention in relation to 
the  phenomenological studies on  relativistic heavy-ion collisions 
\cite{Hirano:2012kj,Asakawa:2015ybt}.

Although the non-perturbative investigations of the EMT at finite $T$ using 
lattice QCD simulations have been very difficult owing to the lack of 
translational and rotational symmetry~\cite{Caracciolo:1990emt,Suzuki:2016ytc}, 
a novel method to construct the EMT on the lattice~\cite{Suzuki:2013gza} 
on the basis of the gradient flow~\cite{Narayanan:2006rf,Luscher:2010iy,Luscher:2011bx}
was recently proposed and was successfully  applied to the equation of states
in pure gauge theory~\cite{Asakawa:2013laa,Kamata:2016any,Kitazawa:2016dsl}%
\footnote{For other recent progress in the construction 
of the EMT on the lattice, see 
Refs.~\cite{Suzuki:2016ytc,Giusti:2010bb,Giusti:2015daa}}.
This study shows that the thermodynamical observables such as the energy density and pressure 
extracted from the expectation values of the EMT (the one-point functions)
agree extremely well with previous high-precision results using the integral 
method~\cite{Boyd:1996bx,Okamoto:1999hi,Borsanyi:2012ve}. 
Also,  the statistics required in the new method is substantially smaller than that in the previous method.
The method is now extended to full QCD simulations at finite $T$~\cite{Makino:2014taa,Taniguchi:2016ofw}.

In the present paper, we report our exploratory studies to extend the 
previous results of   the one-point functions to the 
two-point EMT correlators in SU(3) lattice gauge theory~\cite{Kitazawa:2014uxa}.
The advantages  of such extension are threefold.
First of all, the method allows direct access to the specific heat $c_V$ 
and entropy density $s$ from the EMT correlations.
Secondly, one could explicitly check the conservation law of EMT obtained by the 
gradient flow.  Thirdly, the method will open the new door to the study of 
important transport coefficients such as the shear and bulk viscosities
\cite{Karsch:1986cq,Nakamura:2004sy,Huebner:2008as,Meyer:2011gj,Astrakhantsev:2017nrs}.
We will focus on the first two aspects in this paper.

Let us here summarize the properties of the 
correlators of the EMT, ${\cal T}_{\mu\nu}(x)$, in the Euclidean 
and continuum spacetime,
where $x_{\mu=1,2,3,4} = (\vec{x}, \tau)$ with $0 \le \tau < 1/T$.
We define a dimensionless temporal correlator
of ${\cal T}_{\mu\nu}(x)$ at finite $T$ and at finite volume $V$ as 
\begin{align}
C_{\mu\nu;\rho\sigma}(\tau)
\equiv \frac1{T^5} \int_V d^3 x 
\langle \delta {\cal T}_{\mu\nu}(x) 
\delta {\cal T}_{\rho\sigma}(0)  \rangle ,
\label{eq:C} 
\end{align}
where $\langle \cdot \rangle $ denotes the thermal average.
We have defined 
$\delta {\cal T}_{\mu\nu}(x) 
\equiv {\cal T}_{\mu\nu}(x)  - \langle {\cal T}_{\mu\nu}(x) \rangle$, so that
$C_{\mu\nu;\rho\sigma}(\tau)$ contains only connected contribution.
Owing to the conservation of the EMT in the Euclidean space-time 
($\partial_{\mu} {\cal T}_{\mu \nu}=0$), 
we have 
$\frac{d}{d\tau} \overline{\cal T}_{4\nu} =0$ with 
$\overline{\cal T}_{\mu\nu} \equiv \int_V d^3 x \ {\cal T}_{\mu\nu}(x)$.
For $\tau\ne0$, this leads to
\begin{align}
\frac{d}{d\tau} C_{4\nu;\rho\sigma}(\tau) = 0.  
\label{eq:ddtC}
\end{align}

Since the energy density of the system is represented as
$\varepsilon  =  - \langle \overline{\cal T}_{44} \rangle /V$, 
the specific heat per unit volume 
$c_V$ is given by~\cite{Asakawa:2015ybt}
\begin{align}
\frac{c_V}{T^3} = \frac1{T^3} \frac{d \varepsilon  }{dT} 
= \frac{\langle (\delta \overline{\cal T}_{44})^2 \rangle }{VT^5} 
= C_{44;44}(\tau) ,
\label{eq:C0000}
\end{align}
where Eq.~(\ref{eq:ddtC}) is used in the last equality.
Note that $\tau$ can be taken anywhere in the range $0<\tau<1/T$
owing to the EMT conservation.

Similarly, from the thermodynamic relation for entropy density
$s=dp/dT = (1/V)d \langle \overline{\cal T}_{11} \rangle/dT$
\cite{Landau-Lifshitz},
one obtains
\begin{align}
\frac{s}{T^3} &= \frac1{T^3} \frac{dp}{dT} 
= \frac{\langle \delta \overline{\cal T}_{44}   \delta \overline{\cal T}_{11} \rangle }{VT^5} 
= - C_{44;11}(\tau).
\label{eq:C0011}
\end{align}
Again, $\tau$ can be taken arbitrarily in $0<\tau<1/T$.

Finally, the momentum fluctuation is related to the enthalpy density 
$h$~\cite{Giusti:2010bb,Minami:2012hs},
\begin{align}
\frac{h}{T^4} 
= \frac {\langle (\delta \overline{T}_{41})^2 \rangle }{VT^5}
= -C_{41;41}(\tau).
\label{eq:C0101} 
\end{align}
At zero chemical potential, $h=\varepsilon + p =s T$.

In the present study, we use the EMT operator defined 
through  the gradient flow~\cite{Suzuki:2013gza}.
The gradient flow for Yang-Mills gauge field is defined by 
the differential equation with respect to the
 hypothetical 5-th coordinate $t$~\cite{Luscher:2010iy}
\begin{align}
\frac{d A_\mu(t,x)}{dt} = - g_0^2 
\frac{ \delta S_{\rm YM}(t)}{ \delta A_\mu (t,x)}
= D_\nu G_{\nu\mu}(t,x) ,
\label{eq:GF}
\end{align}
with the Yang-Mills action $S_{\rm YM}(t)$ and 
the field strength $G_{\mu\nu}(t,x)$ composed of the 
transformed field $A_\mu(t,x)$.
The flow time $t$ has a dimension of inverse mass squared.
The initial condition at $t=0$ is taken for the field in the
conventional gauge theory; $A_\mu(0,x)=A_\mu (x)$.
The gradient flow for positive $t$ acts as the smearing of the gauge 
field with the smearing radius $\sqrt{8t}$
~\cite{Luscher:2010iy}.
The EMT operator is then defined  as 
\begin{eqnarray}
  {\cal T}_{\mu\nu}(x)  &= &\lim_{t\to0} {\cal T}_{\mu\nu}(t,x) , \label{eq:t} \\ 
  {\cal T}_{\mu\nu}(t,x)  &= &
  \frac{U_{\mu\nu}(t,x)}{\alpha_U(t)}
  +\frac{\delta_{\mu\nu}}{4\alpha_E(t)}
  \left[E(t,x)-\left\langle E(t,x)\right\rangle_0 \right] , \nonumber \\ 
\label{eq:T}
\end{eqnarray}
where the dimension-four gauge-invariant operators on the right hand side are
given by \cite{Suzuki:2013gza}
\begin{align}
  E(t,x) &= \frac14 G_{\mu\nu}^a(t,x)G_{\mu\nu}^a(t,x),
  \label{eq:E} 
  \\
  U_{\mu\nu}(t,x) &= G_{\mu\rho}^a (t,x)G_{\nu\rho}^a (t,x)
  - \delta_{\mu\nu} E(t,x), 
  \label{eq:U}
\end{align}
while $\langle E(t,x) \rangle_0$ in Eq.~(\ref{eq:T})  is the vacuum expectation value
of $E(t,x)$, which is introduced so that 
$\langle {\cal T}_{\mu\nu}(x) \rangle$ vanishes in the vacuum.
The coefficients $\alpha_U(t)$ and $\alpha_E(t)$ have been calculated
perturbatively in Ref.~\cite{Suzuki:2013gza} for small $t$: 
Their explicit forms in the $\overline{\mathrm{MS}}$ scheme are 
given in Ref.~\cite{Kitazawa:2016dsl}.

Although Eqs.~(\ref{eq:t}) and (\ref{eq:T}) are exact 
in the continuum spacetime,
special care is required in lattice gauge theory with finite lattice spacing $a$: 
The flow time should satisfy  $\sqrt{8t} \gtrsim a$ to suppress the lattice discretization 
effects.  It has been shown for the thermal average of the EMT that
there exists indeed a range of $t$ for sufficiently small $a$,  so that
the lattice data allow reliable 
extrapolation to  $t=0$ to obtain $\varepsilon$ and 
$p$~\cite{Asakawa:2013laa,Kamata:2016any,Kitazawa:2016dsl}.

To analyze  the two-point EMT correlations with Eq.~(\ref{eq:T}), 
we have an extra condition that
the distance between the two smeared operators $\tau$ 
in temporal direction is well separated (with the temporal periodicity) 
to avoid their overlap.  Because the smearing length along temporal direction is 
$\sqrt{2t}$, the necessary conditions read
$a \lesssim  2 \sqrt{2t} \lesssim  \tau  \le  1/(2T) $,
or equivalently, in terms of the dimensionless quantities, as
\begin{eqnarray}
\frac{1}{N_{\tau}} \lesssim \sqrt{8  tT^2} \lesssim  \tau T \le \frac{1}{2}
\label{eq:tau>t}
\end{eqnarray}
with $N_{\tau}=(aT)^{-1}$ being the temporal lattice size.

\begin{figure*}
  \centering
  \includegraphics[width=0.313\textwidth,clip]{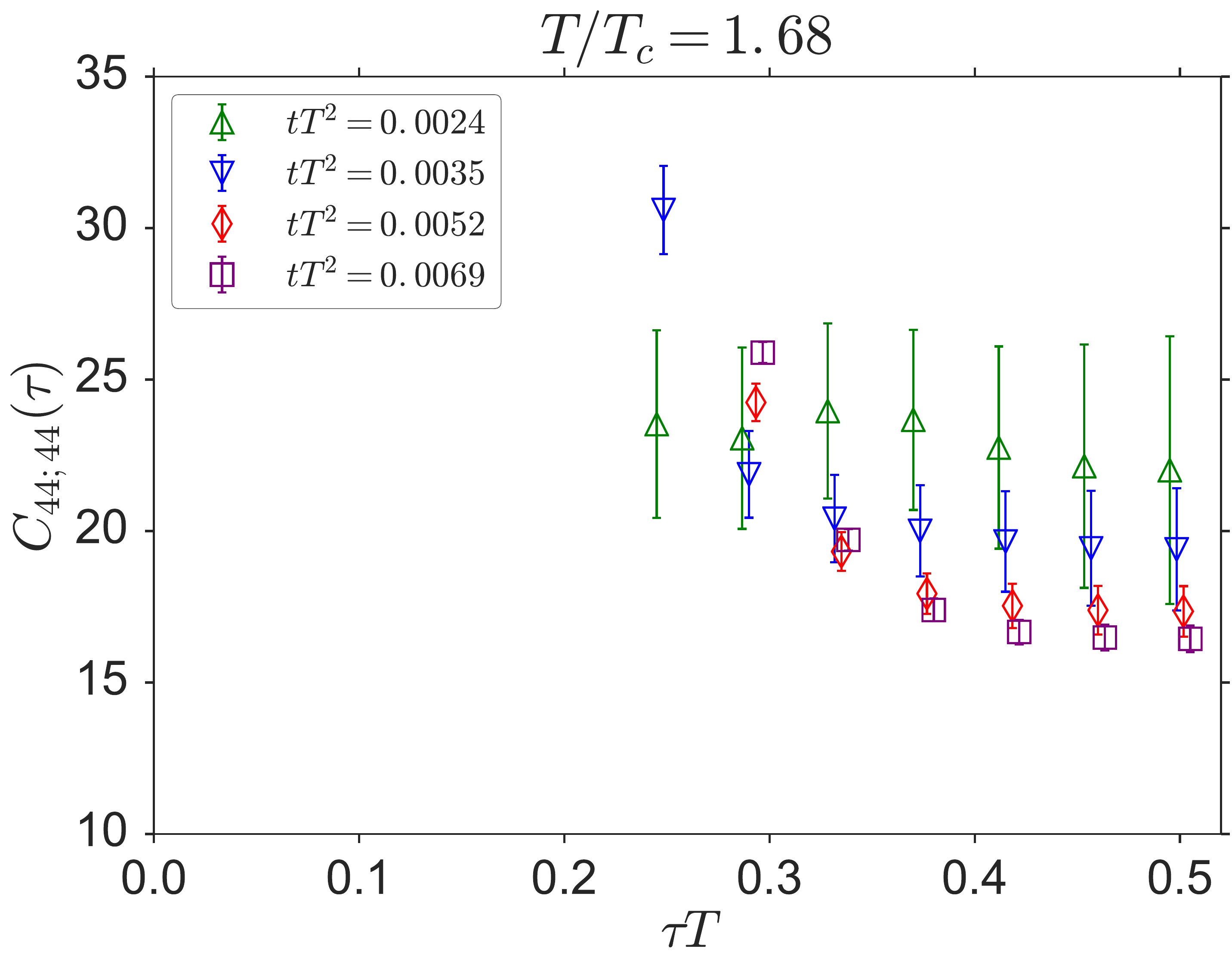}
  \includegraphics[width=0.32\textwidth,clip]{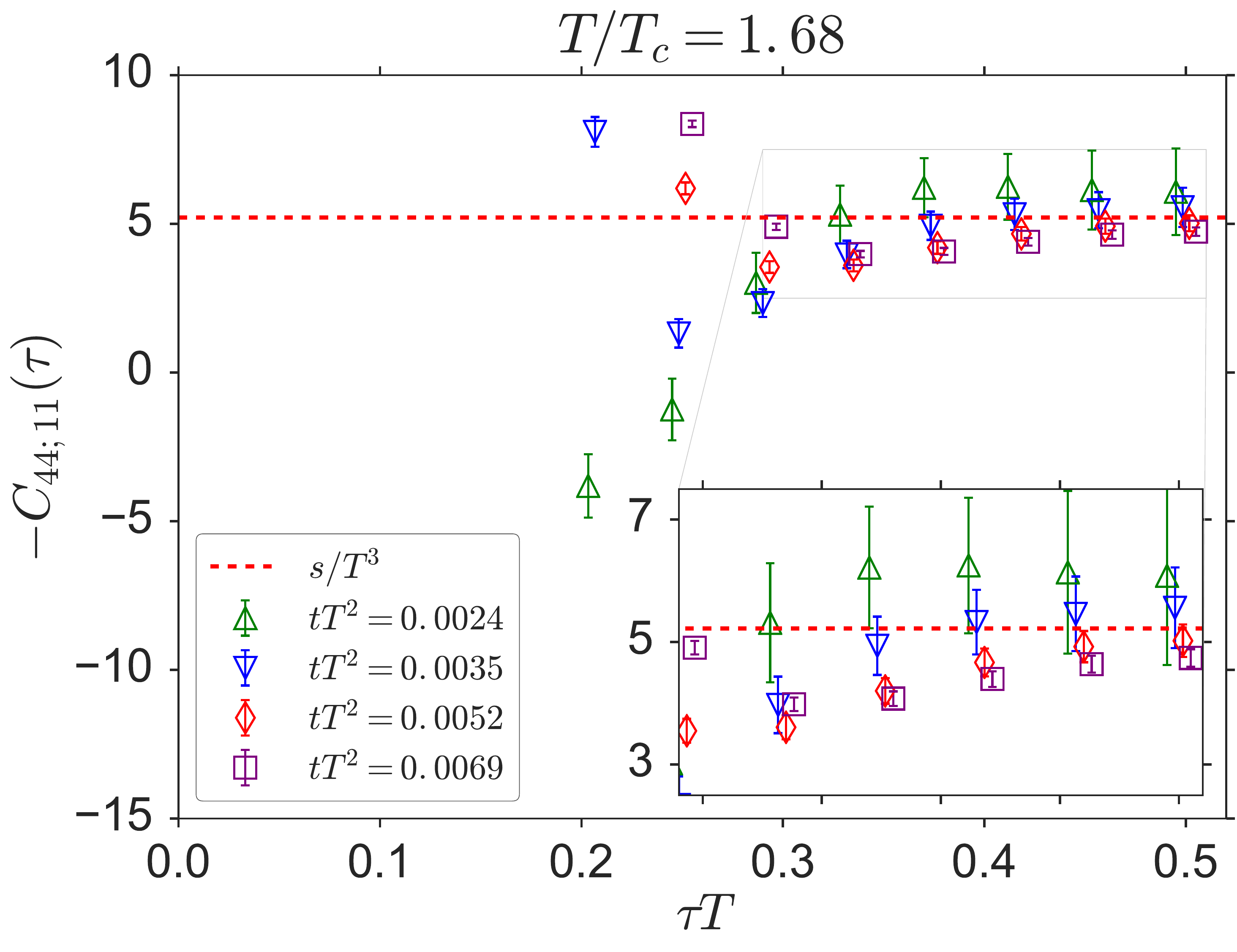}
  \includegraphics[width=0.32\textwidth,clip]{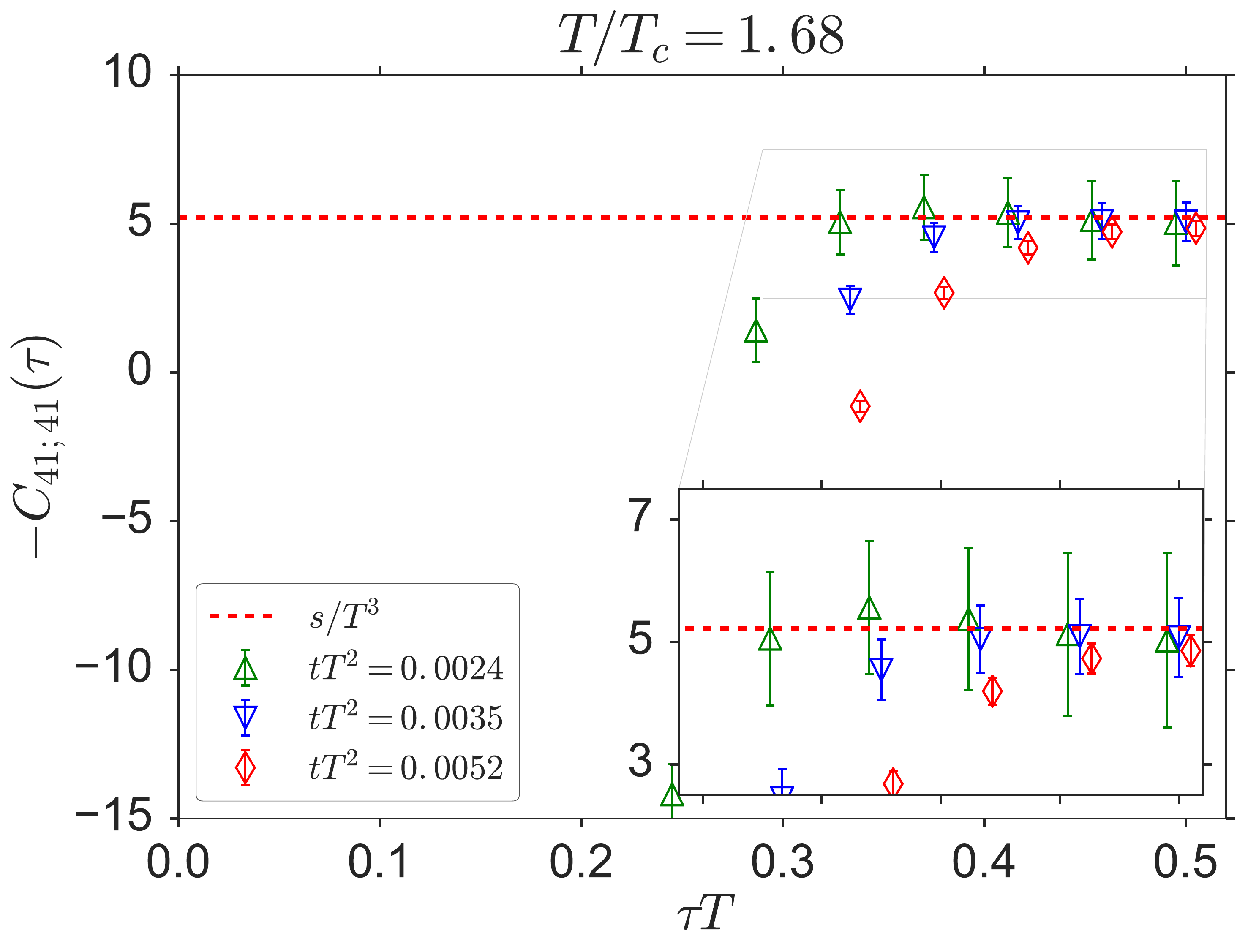}
  
  \includegraphics[width=0.313\textwidth,clip]{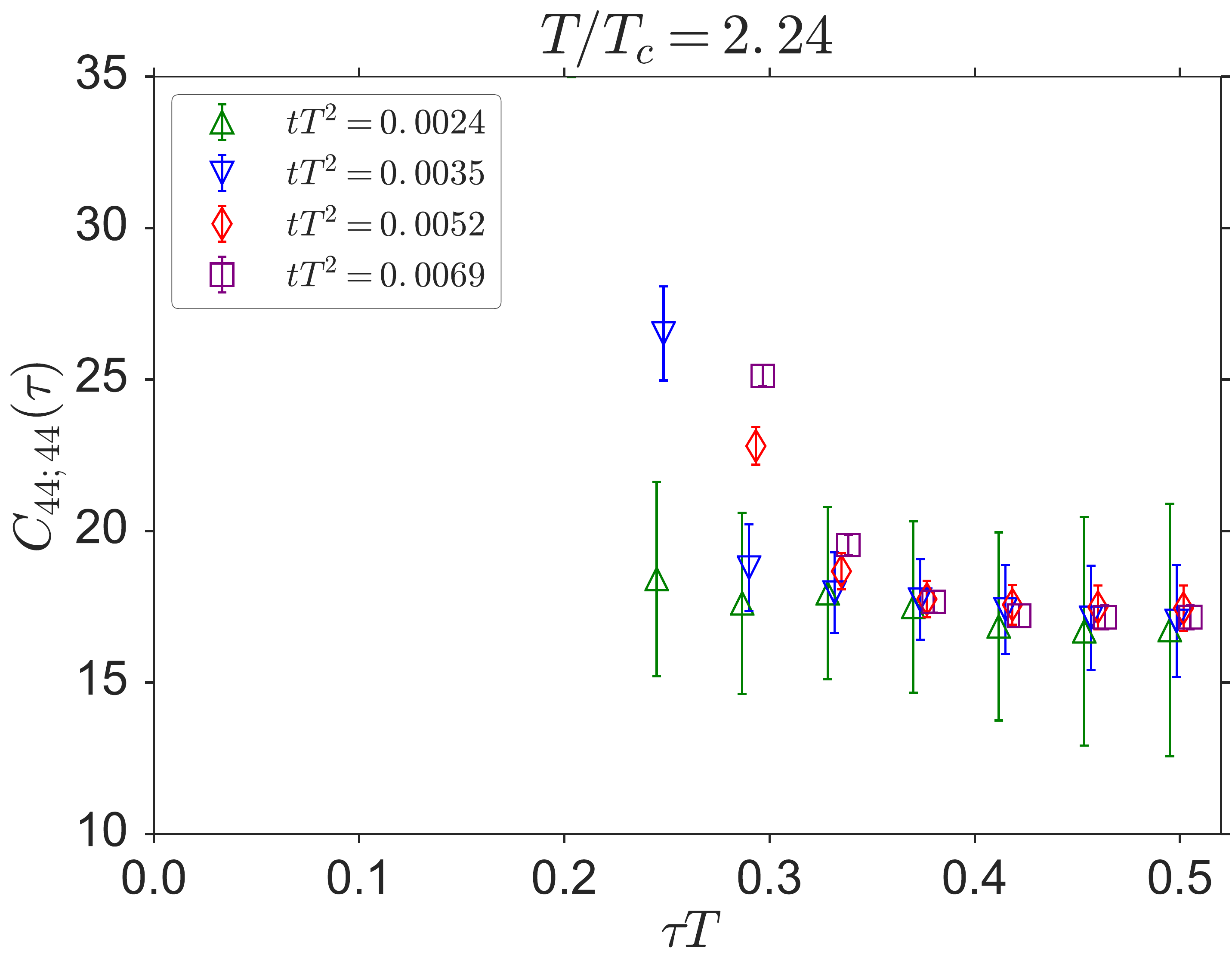}
  \includegraphics[width=0.32\textwidth,clip]{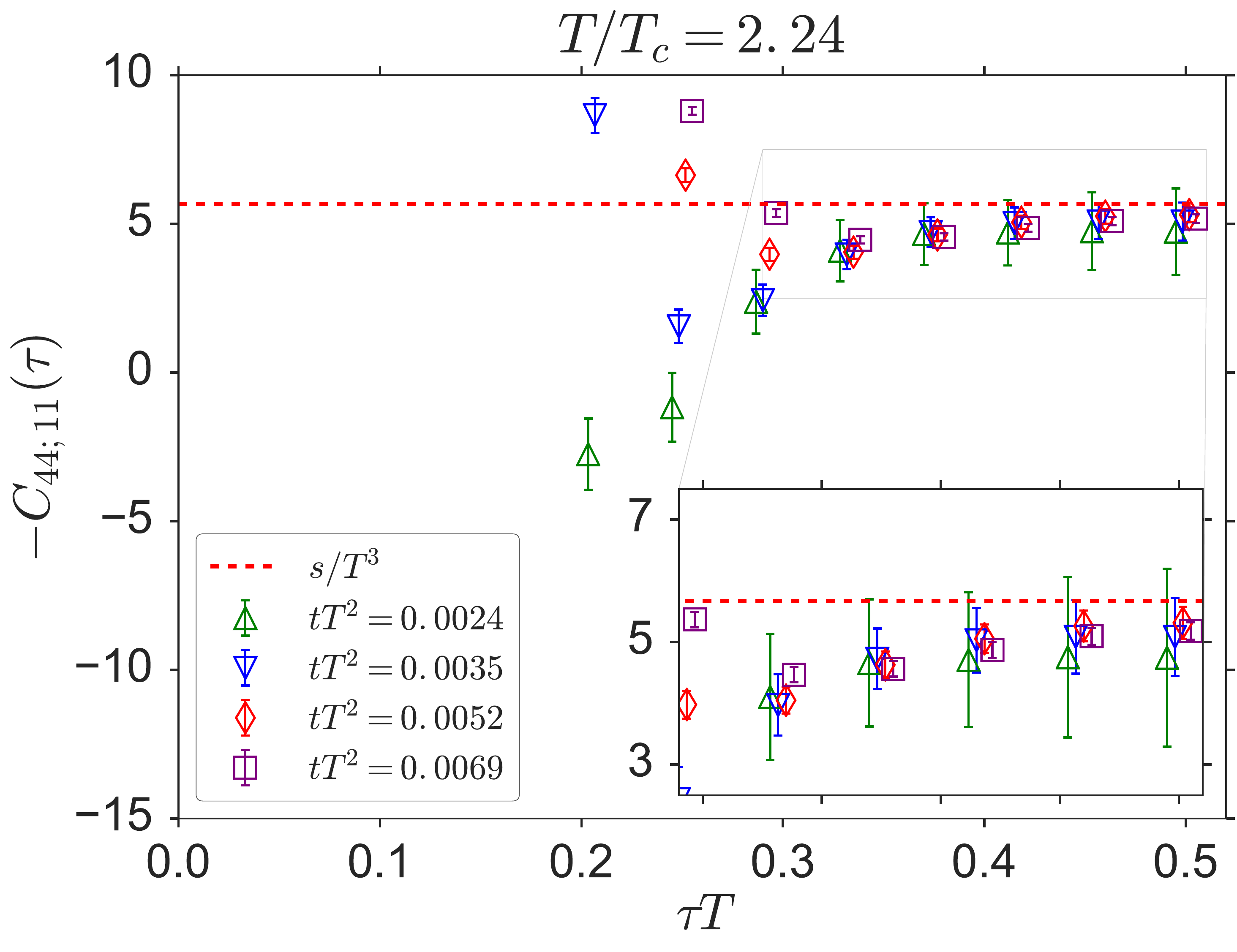}
  \includegraphics[width=0.32\textwidth,clip]{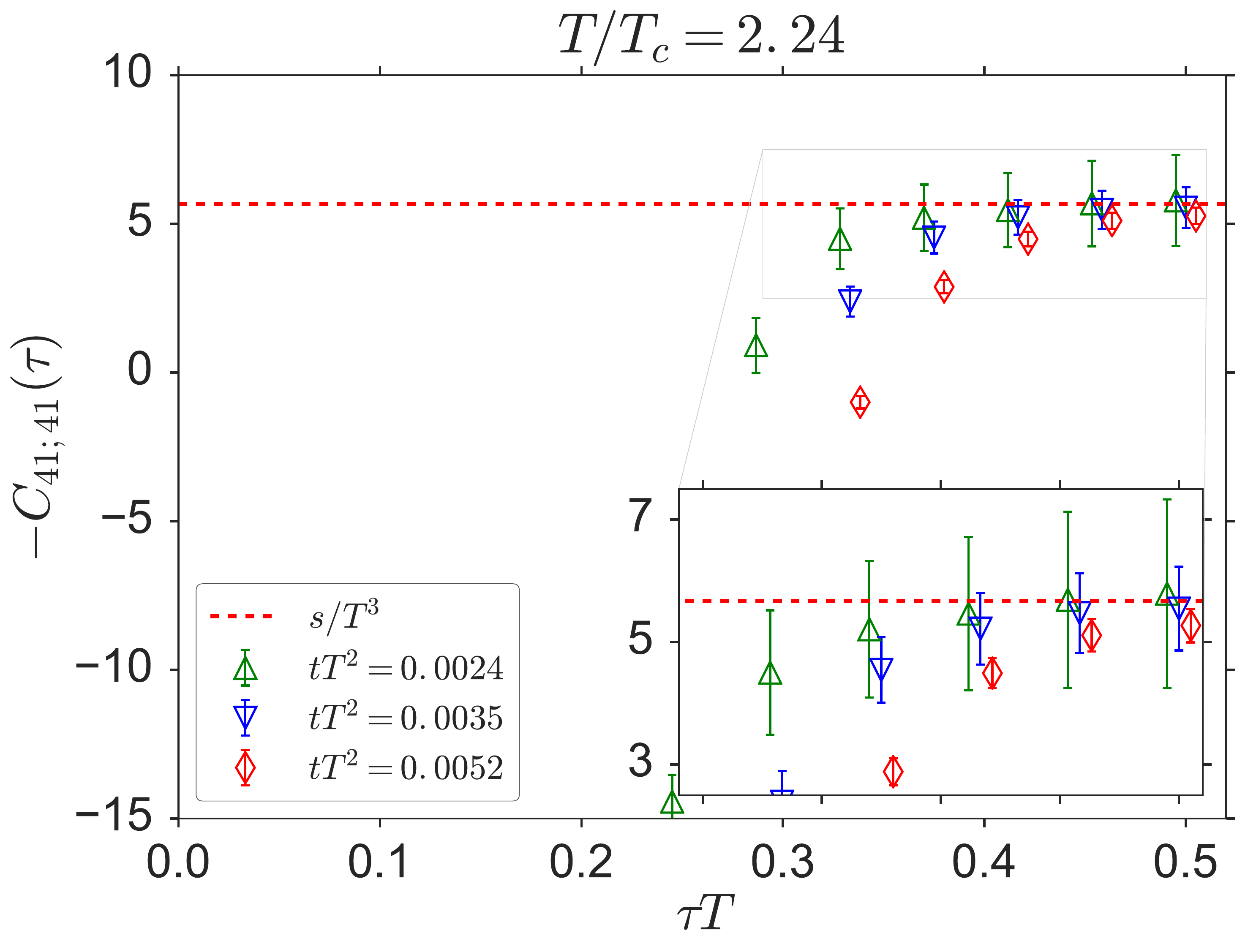}
\caption{
  Correlators $C_{44;44}(\tau)$ (left),
  $C_{44;11}(\tau)$ (middle), and $C_{41,41}(\tau)$ (right)
  for several values of flow time $t$
  for $N_\tau=24$.
  The red dashed lines show $s/T^3=(\varepsilon+p)/T^4$ obtained from
  the one-point function of the EMT with the same gauge configurations.
}
\label{fig:cor}
\end{figure*}

\begin{table}
\centering
\begin{tabular}{cccc}
\multicolumn{4}{c}{$T/T_{\rm c}=1.68$} \\
\hline \hline
$N_s$ & $N_\tau$ & $\beta$ & $N_{\rm conf}$ \\
\hline
96 & 24 & 7.265 & 200,000 \\
64 & 16 & 6.941 & 180,000 \\
48 & 12 & 6.719 & 180,000 \\
\hline \hline
\end{tabular}
\hspace{5mm}
\begin{tabular}{cccc}
\multicolumn{4}{c}{$T/T_{\rm c}=2.24$} \\
\hline \hline
$N_s$ & $N_\tau$ & $\beta$ & $N_{\rm conf}$ \\
\hline
96 & 24 & 7.500 & 200,000 \\
64 & 16 & 7.170 & 180,000 \\
48 & 12 & 6.943 & 180,000 \\
\hline \hline
\end{tabular}
\caption{
Simulation parameters on the lattice: $N_s$, $N_{\tau}$, $\beta=6/g_0^2$,
and $N_{\rm conf}$ are
spatial lattice size, temporal lattice size, the bare coupling constant, and
 the total number of gauge configurations, respectively.
}
\label{table:param}
\end{table}

In our numerical studies, we consider SU(3) Yang-Mills theory
on four-dimensional Euclidean lattice and 
employ the Wilson gauge action under the periodic boundary condition.
Gauge configurations are generated by the same procedure as in 
Ref.~\cite{Kitazawa:2016dsl}, but each measurement is separated 
by $50$ sweeps.
Statistical errors are estimated by the jackknife method
with $100$ jackknife samples.
On the right hand side of the flow equation Eq.~(\ref{eq:GF}),
the Wilson gauge action is used for  $S_{\rm YM}(t)$, while 
the operators in Eqs.~(\ref{eq:E}) and (\ref{eq:U}) are constructed from 
$G^a_{\mu\nu}(t,x)$ defined by the clover-type representation. 

We study two cases above the deconfinement transition,
$T/T_{\rm c}=1.68$ and $2.24$, 
with three different lattice volumes $N_s^3\times N_\tau$ 
with a fixed aspect ratio $N_s/N_\tau=4$.
The values of $\beta=6/g_0^2$ corresponding to each set of
$T/T_{\rm c}$ and $N_{\tau}$ are
obtained from Refs.~\cite{Asakawa:2015vta,Kitazawa:2016dsl}.
The resultant simulation parameters are summarized 
in Table~\ref{table:param}.

Shown in  Fig.~\ref{fig:cor} are the $\tau$ dependences of 
$C_{44;44}(\tau)$, $C_{44;11}(\tau)$, and $C_{41;41}(\tau)$ 
for $T/T_{\rm c}=1.68$ (upper panels)  and $T/T_{\rm c}=2.24$ (lower panels)
for  $N_\tau=24$ with typical values of $tT^2$
between the upper and lower bounds in Eq.~(\ref{eq:tau>t}).
From the overall behavior of the lattice data in these figures, one finds two key features: 
(i) As $t$ decreases, the data start to show the  plateau structure for 
$\tau T \gtrsim 0.3$.  
(ii) As $t$ decreases, the statistical errors become larger.
The feature (i) is a signature of the EMT conservation Eq.~(\ref{eq:ddtC})
for large $\tau$, where the smeared EMT operators do not overlap with each other.
The feature (ii) is due to the fact that the 
gauge fields are rough (smooth) for small (large) $t$.  

Shown by the red dashed lines 
in Fig.~\ref{fig:cor} together with  $C_{44;11}(\tau)$ and $C_{41;41}(\tau)$ are   $s/T^3=(\varepsilon+p)/T^4$ obtained by 
the one-point function of EMT ($\langle \overline{\cal T}_{44} \rangle$ and $\langle \overline{\cal T}_{11} \rangle$) 
using the 
method in Ref.~\cite{Kitazawa:2016dsl} with the same configurations.
This agreement of the results of $s/T^3$ between the one-point function
and the two-point functions, as it should be for Eqs.~(\ref{eq:C0011}) and (\ref{eq:C0101}),
at large $\tau$ and small $t$ indicates
an internal consistency of the present method.

\begin{figure*}
  \centering
  \includegraphics[width=0.32\textwidth,clip]{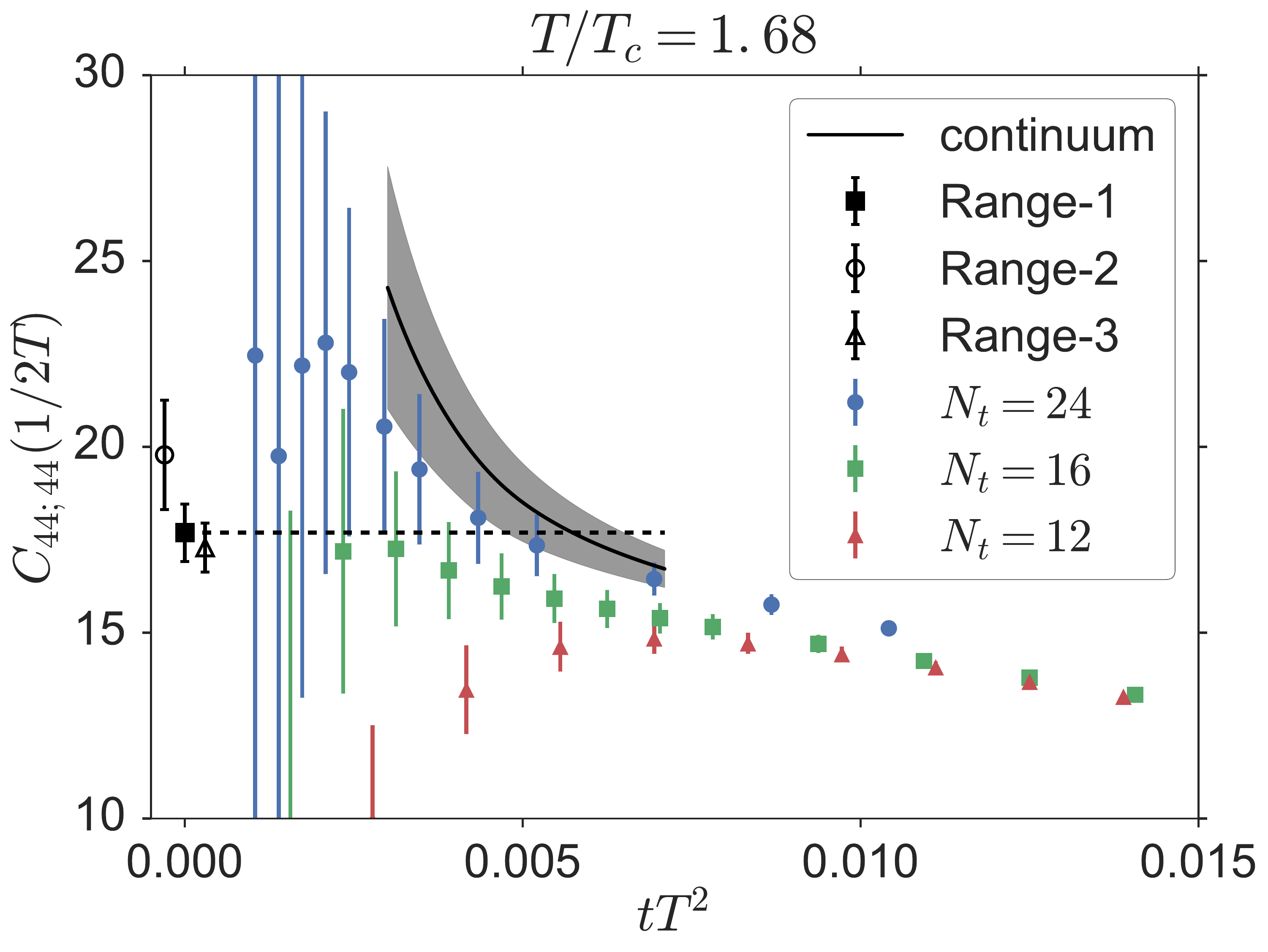}
  \includegraphics[width=0.32\textwidth,clip]{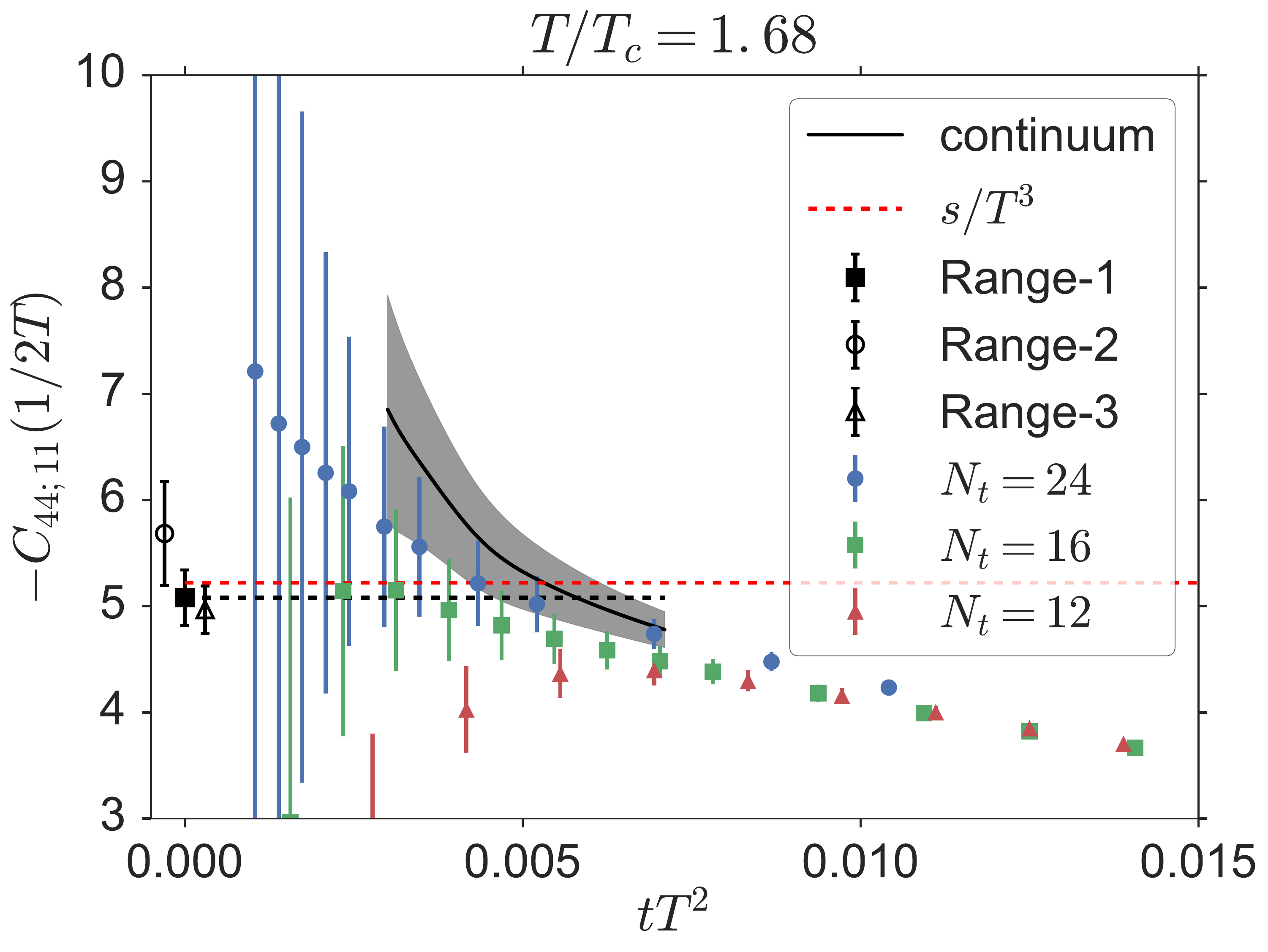}
  \includegraphics[width=0.32\textwidth,clip]{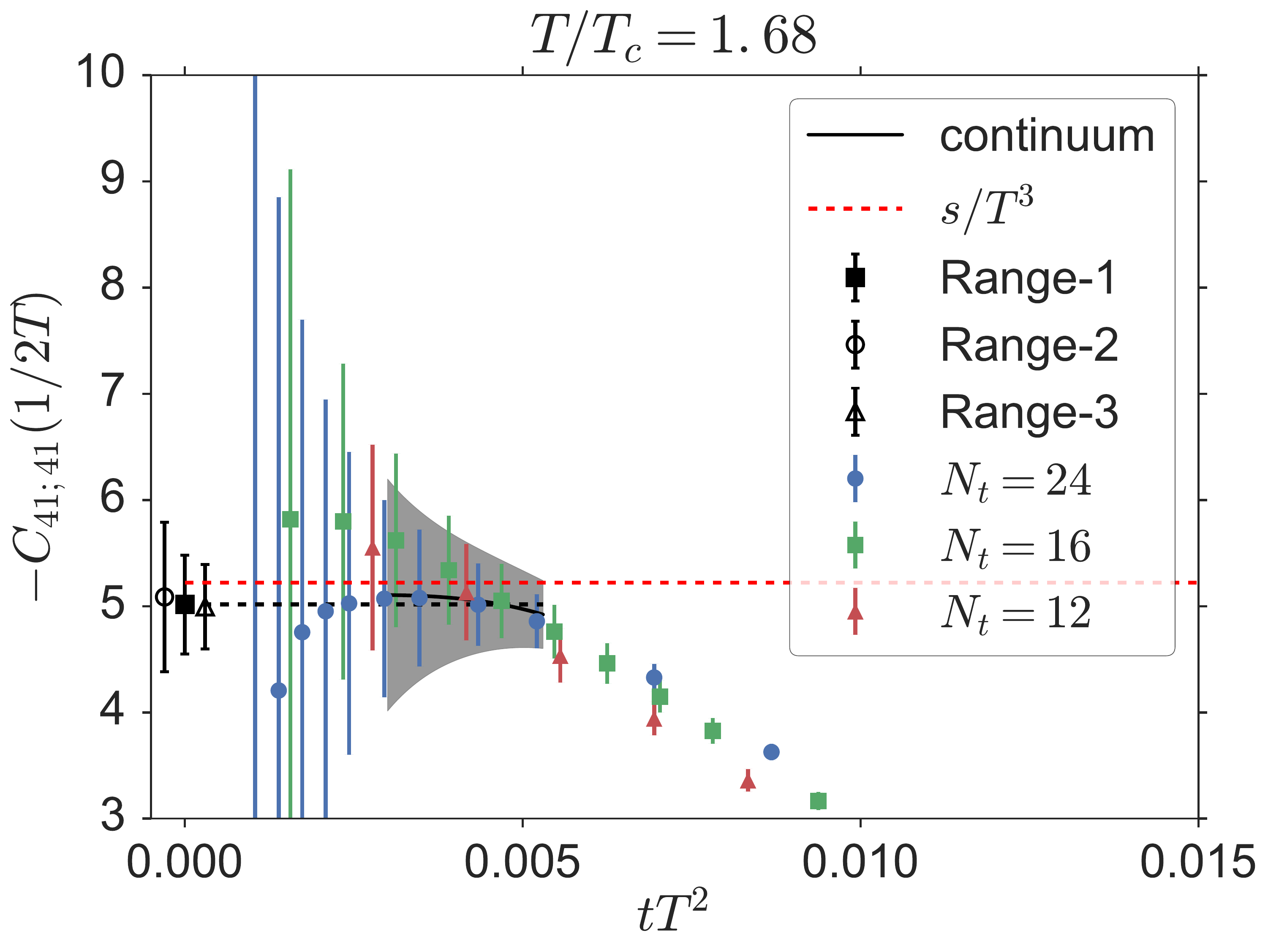}
  \includegraphics[width=0.32\textwidth,clip]{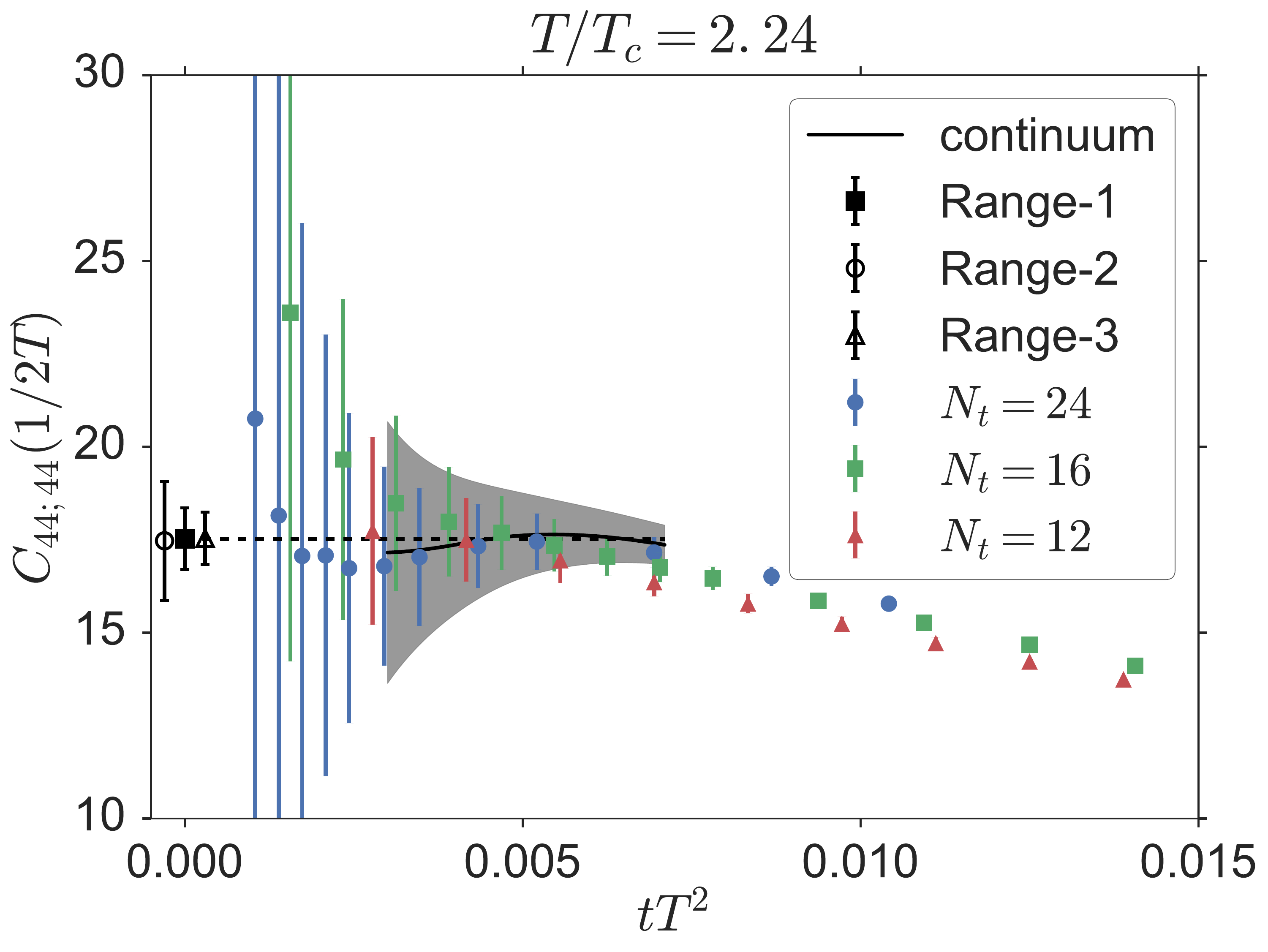}
  \includegraphics[width=0.32\textwidth,clip]{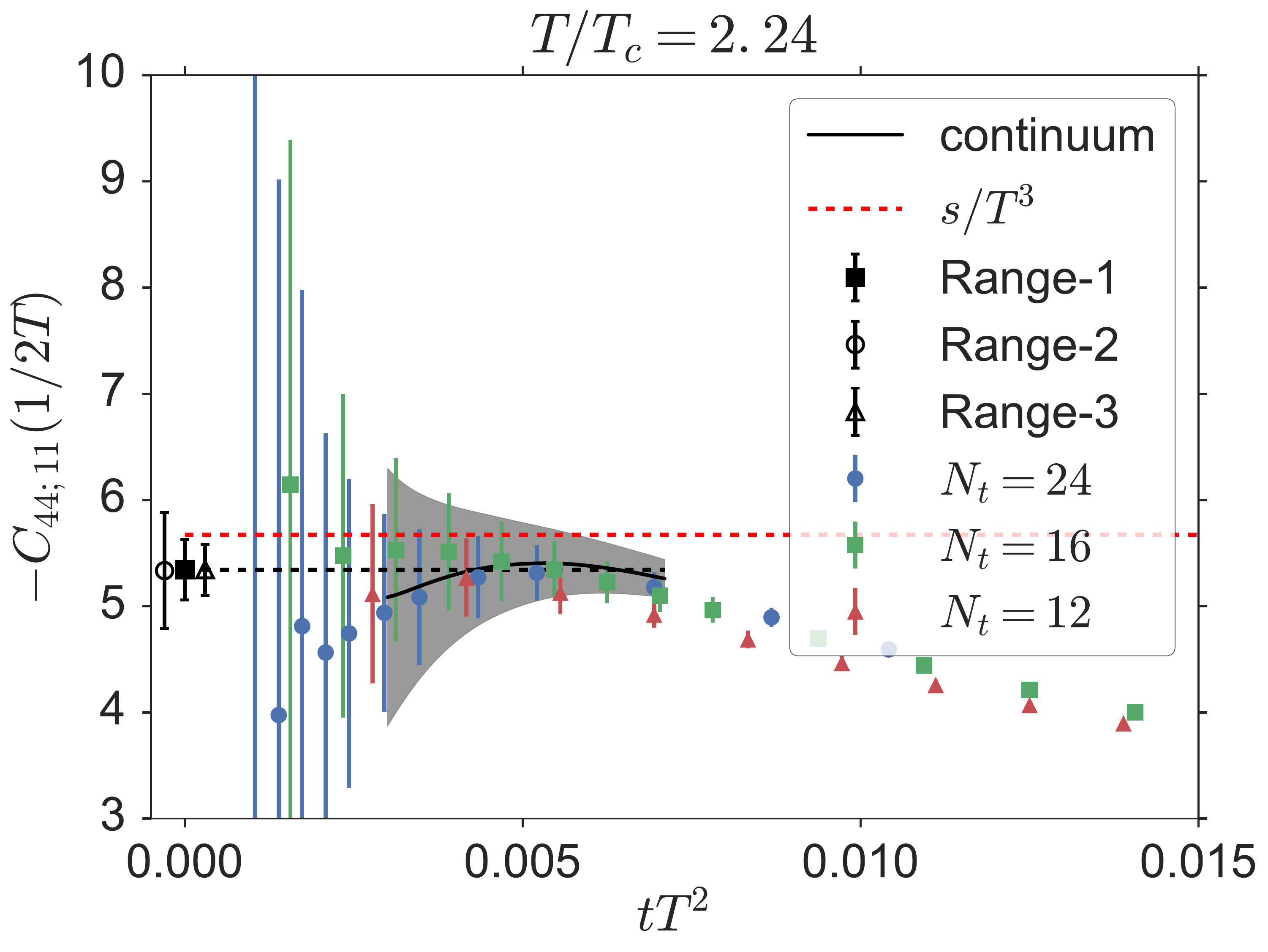}
  \includegraphics[width=0.32\textwidth,clip]{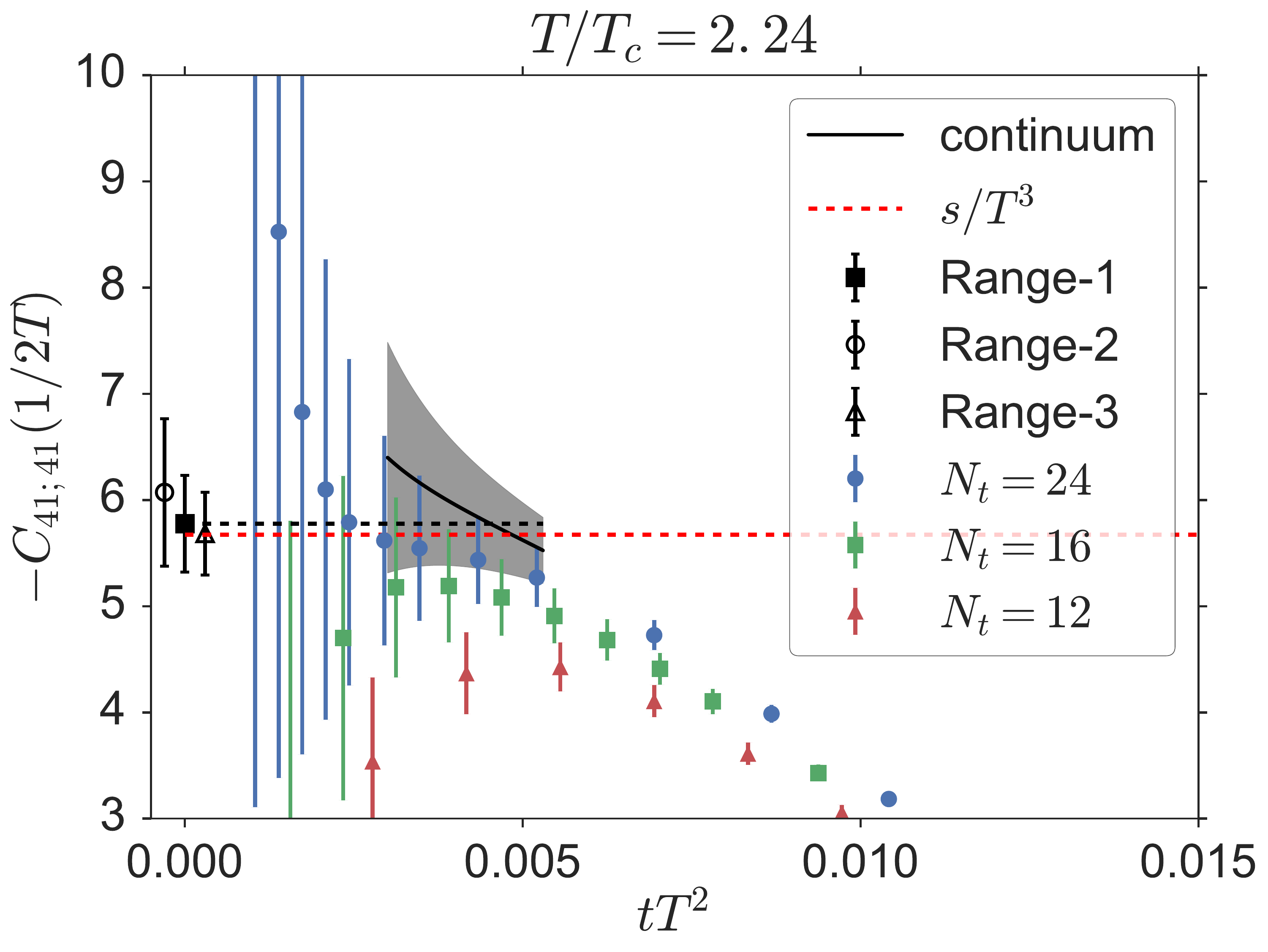}
\caption{
  $t$ dependences of the midpoint correlators 
  $C_{44,44}(\tau_m)$ (left), 
  $C_{44,11}(\tau_m)$ (middle), and
  $C_{41,41}(\tau_m)$ (right)
  for $T/T_{\rm c}=1.68$ and $2.24$.
  The black lines show the results of continuum extrapolation
  using $N_\tau = 12, 16$, and $24$.
}
\label{fig:mid}
\end{figure*}

To  take the continuum limit $a\rightarrow 0$ followed by an
extrapolation $t\rightarrow 0$,
we show, in Fig.~\ref{fig:mid}, the $t$ dependence of the correlators for 
different lattice spacings, $a=1/(N_{\tau}T)$. Here we choose the maximum possible 
separation, $\tau= \tau_m=1/(2T)$, to minimize  the overlap of the EMT operators.
The  continuum extrapolation is carried out 
by using the data at $N_\tau=12$, $16$, and $24$ for each  $t \in [t_{\rm min}, t_{\rm max}] $
with an ansatz
$C_{\mu \nu; \rho \sigma} (\tau_m) |_{_{\rm lat}} =  
C_{\mu \nu; \rho \sigma} (\tau_m)|_{_{\rm cont}} + O(a^2)$ expected from 
perturbation theory.
Here $t_{\rm min,~max}$ is chosen in such a way that
$C_{4\nu;\rho\sigma}(\tau_m)/C_{4\nu;\rho\sigma}(\tau_m-a) = 1 $ 
within the statistical errors: This procedure excludes the small $t$ region where large 
lattice discretization errors arise, as well as  
the large $t$ region where the systematic
errors from the overlap of EMT operators and  
the contribution of
higher dimensional operators other than Eqs.~(\ref{eq:E}) and (\ref{eq:U}) 
are not negligible~\cite{Kitazawa:2016dsl}.
The resulting ranges in the dimensionless combination are
$(tT^2)_{\rm min} = 0.003$ and  $(tT^2)_{\rm \max} = 0.007$  for $C_{44;44}(\tau_m)$, and
$(tT^2)_{\rm min} = 0.003$ and  $(tT^2)_{\rm max} = 0.006$  for $C_{44;11}(\tau_m)$ and
$C_{41;41}(\tau_m)$. 

The results of the continuum extrapolation 
are shown by the black lines with the error represented by the gray band
in Fig.~\ref{fig:mid}. At  the level of error bars in the present exploratory  study, we do not 
have enough resolution to reliably extract the $O(t)$ contribution in 
$C_{\mu \nu; \rho \sigma} (\tau_m)$~\cite{Kitazawa:2016dsl},
so that  we take the $t\to0$ extrapolation by a constant fit in the interval
$ [t_{\rm min}, t_{\rm max}] $, which is called Range-1.  The final results after
the double extrapolation ($t\to0$ after $a\to0$) 
are shown by the filled squares in  Fig.~\ref{fig:mid}.  To estimate the systematic
errors from this constant fit, we choose the Range-2 (the first half of Range-1) 
and Range-3 (the latter half of Range-1); the results are shown by
open circles and open triangles, respectively. 
The red dashed lines in the middle and right panels in Fig.~\ref{fig:mid} are 
$s/T^3$ obtained from the one-point function of EMT.

\begin{table}
\centering

\begin{tabular}{|c|c|c|c|c|}
\multicolumn{5}{c}{$s/T^3$}  \\
 \hline
$T/T_{\rm c}$ & $C_{44;11}(\tau_m)$ & $C_{41;41}(\tau_m)$  & $ \langle {\cal T}_{\mu \nu} \rangle$  & ideal gas  \\
\hline \hline
1.68  &  5.08(26)$(^{+60}_{-11}$)   & 5.02(47)$(^{+7}_{-2})$ & 5.222(10)(24) & 7.02\\
\hline
2.24  & 5.34(28)$(^{+0}_{-0}$) & 5.78(46) $(^{+29}_{-10})$ & 5.675(10)(24)  & 7.02 \\
\hline
\end{tabular}

\caption{
  Values of $s/T^3$ obtained from 
  Eqs.~(\ref{eq:C0011}) and (\ref{eq:C0101}) together with those
  obtained from the one-point function of EMT.
  The ideal gas limit for massless gluons
  is also shown for comparison.  The first (second) parenthesis shows statistical (systematic) error.   
  The systematic error for the one-point function originates from 
  the 1\%  uncertainty of $\Lambda_{\overline{\rm MS}}$~\cite{Kitazawa:2016dsl}.
}
\label{table:result-s}
\end{table}

\begin{table}
\centering

\begin{tabular}{|c|c|c|c|c|}
  \multicolumn{5}{c}{$c_V/T^3$}  \\
  \hline
  $T/T_{\rm c}$ & $C_{44;44}(\tau_m)$   & Ref.~\cite{Gavai:2004se} &  Ref.~\cite{Borsanyi:2012ve}  & ideal gas\\
  \hline \hline
  1.68  & 17.7(8)$(^{+2.1}_{-0.4})$  & 22.8(7)$^*$ & 17.7  & 21.06 \\
  \hline
  2.24 & 17.5(8)$(^{+0}_{-0.1})$  &  17.9(7)$^{**}$  &   18.2 & 21.06\\
  \hline 
\end{tabular}

\caption{
  Values of  $c_V/T^3$ obtained by
  Eq.~(\ref{eq:C0000}) as well as those obtained directly from the 
  differential  method \cite{Gavai:2004se} and those calculated  indirectly from $\varepsilon(T)$ in the 
  integral method \cite{Borsanyi:2012ve}.  The ideal gas limit for massless gluons
  is also shown for comparison. The error bars 
  are estimated in
  the same way as $s/T^4$ in Table \ref{table:result-s}.
  The  symbol $*$ ($**$)  indicates that the numbers are for $T/T_c=1.5~(2.0)$.
  The error bars of $c_V/T^3$  in the column Ref.  \cite{Borsanyi:2012ve} would be a few \% level.
}
\label{table:result-c}
\end{table}

Shown in Table~\ref{table:result-s} are the numerical results of $s/T^3$
obtained in the present analysis of EMT correlators.  
Within the statistical and systematic error bars, the results of the two 
different correlators agree with each other, and they agree to the results
of the one-point function of EMT.  Also the central value of $s/T^3$  
in our analysis increases as $T$ and also much less than the ideal gas value,
which captures the essential feature expected from strongly interacting
gluon plasma above $T_{\rm c}$.
 
Shown in Table~\ref{table:result-c} are the numerical results of $c_V/T^3$
obtained in the present analysis of EMT correlators.  
Our results agree  quantitatively with the numbers extracted from the 
recent high-precision study of the energy density in the integral method \cite{Borsanyi:2012ve}
and  qualitatively with the numbers obtained in the 
differential method \cite{Gavai:2004se}.  Our specific heat is about 20\%
smaller than the ideal gas value, which also indicates the strong coupling feature of the 
system.

In summary, we have investigated the two-point  EMT correlators 
in SU(3) Yang-Mills theory at finite temperature ($T/T_{\rm c}=2.24$ and $1.68$)
using the method of gradient flow with the flow time $t$.
The correlators $C_{4\nu;\rho\sigma}(\tau)$  approach constant values 
for sufficiently large $\tau$ and small $t$.  This is an indication that the
conservation of the EMT is realized in the gradient flow  as long as the two EMT operators do not have
overlap with each other.   
By taking the double limit ($t \rightarrow 0$ after $a \rightarrow 0$)
using the data for $N_{\tau}=12$, $16$, and $24$, we found that 
the entropy density ($s$) obtained from the two-point EMT correlators 
($C_{44;11}(\tau_m)$ and $C_{41;41}(\tau_m)$) reproduces the 
high precision result previously  obtained from the one-point function.
Also, we found that the specific heat ($c_V$) can be 
determined in 5-10\% accuracy from
the two-point EMT correlator ($C_{44;44}(\tau_m)$) 
even with relatively low statistics.
Now that we have confirmed that thermodynamical quantities
are obtained accurately with two-point EMT correlators with
the gradient flow, it is within reach to 
investigate
transport coefficients with two-point EMT correlations as well.

Although we focused on SU(3) Yang-Mills theory in this study,
the same analysis can be also performed in full QCD 
\cite{Makino:2014taa,Taniguchi:2016ofw}.  A preliminary study along 
this line is reported in Ref.~\cite{Taniguchi:2017}.

The authors thank E.~Itou and H.~Suzuki for discussions 
in the early stage.
Numerical simulation was carried out on IBM System Blue Gene Solution 
at KEK under its Large-Scale Simulation Program (Nos.~13/14-20, 14/15-08, 15/16-15, 16/17-07). 
This work was supported by JSPS KAKENHI Grant Numbers JP17K05442 and
25287066. 
TH were partially supported by the RIKEN iTHES Project and iTHEMS Program.
TH is  grateful to the Aspen Center for Physics, 
supported by NSF Grant PHY1607611, where part of this research was done.

\end{document}